\documentclass[twocolumn,showpacs,amsmath,amssymb,prl,superscriptaddress,floatfix]{revtex4}

\usepackage{graphicx,color}
\usepackage{mathptmx, textcomp}
\usepackage[latin1]{inputenc}
\usepackage{braket,amsfonts}
\usepackage{ulem}

\newcommand{\g} {g^{(3)}}
\newcommand{\K} {K^{(3)}}
\newcommand{\scat} { a_\text{3D} }

\newcommand{\Kg} { K^{(3)} g^{(3)} }
\newcommand{\stext}[1] {{\text{\scriptsize #1}}}
\begin{document}

\title{Three-body correlation functions and recombination rates for bosons in three and one dimensions}

\author{E. Haller}
\affiliation{Institut f\"ur Experimentalphysik und Zentrum f\"ur Quantenphysik, Universit\"at Innsbruck, 6020 Innsbruck, Austria}
\affiliation{Institut f\"ur Quantenoptik und Quanteninformation, \"Osterreichische Akademie der Wissenschaften, 6020 Innsbruck, Austria}
\author{M. Rabie}
\author{M.J. Mark}
\author{J.G. Danzl}
\author{R. Hart}
\author{K. Lauber}
\affiliation{Institut f\"ur Experimentalphysik und Zentrum f\"ur Quantenphysik, Universit\"at Innsbruck, 6020 Innsbruck, Austria}
\author{G. Pupillo}
\affiliation{Institut f\"ur Quantenoptik und Quanteninformation,
\"Osterreichische Akademie der Wissenschaften, 6020 Innsbruck, Austria}
\affiliation{Institut f\"ur Theoretische Physik, Universit\"at Innsbruck, 6020 Innsbruck, Austria}
\author{H.-C. N\"agerl}
\affiliation{Institut f\"ur Experimentalphysik und Zentrum f\"ur Quantenphysik, Universit\"at Innsbruck, 6020 Innsbruck, Austria}

\date{\today}


\begin{abstract}
We investigate local three-body correlations for bosonic particles in three  and one dimensions as a function of the interaction strength. The three-body correlation function $\g$ is determined by measuring the three-body recombination rate in an ultracold gas of Cs atoms. In three dimensions, we measure the dependence of $\g$ on the gas parameter in a BEC, finding good agreement with the theoretical prediction accounting for beyond-mean-field effects. In one dimension, we observe a reduction of $\g$ by several orders of magnitude upon increasing interactions from the weakly interacting BEC to the strongly interacting Tonks-Girardeau regime, in good agreement with predictions from the Lieb-Liniger model for all strengths of interaction.
\end{abstract}

\maketitle

Correlation functions reflect the non-classical nature of quantum many-body systems.  They may be used to characterize the latter when quantities such as temperature, density, dimensionality, and particle statistics are varied in experiments. It is particularly instructive to monitor a system's correlation functions as the strength of particle interactions is tuned from weak to strong. A paradigm is given by an ensemble of bosons in one-dimensional (1D) geometry with contact interactions \cite{Cazalilla2011}: For weak repulsive interactions, in the zero-temperature limit, the system is a quasicondensate with essentially flat particle correlation functions in position space to all orders. For strong repulsive interactions, the bosons avoid each other, leading to loss of coherence and strong increase of local correlations. In the context of ultracold atomic gases, with exquisite control over temperature, density, and dimensionality \cite{Bloch2008}, tuning of interactions is enabled by Feshbach resonances \cite{Chin2010}. Local two- and three-body correlations in atomic many-body systems can be probed e.g.~in measurements of photoassociation rates \cite{Kinoshita2005} and of three-body recombination processes \cite{Burt1997,Tolra2004}, respectively. Non-local two-body correlations for atomic matter waves have been measured in atom counting \cite{Yasuda1996,Oettl2005,Schellekens2005,Jeltes2007}, noise-correlation \cite{Greiner2005,Foelling2005,Rom2006}, and in-situ imaging \cite{Jacqmin2011} experiments. Recently, also non-local three-body correlations have become accessible in experiments \cite{Armijo2010,Hodgman2011}.

Recombination processes are sensitive to the properties of the many-body wave  function at short distances. In particular, the process of three-body recombination, in which three particles collide inelastically to form a dimer, is directly connected to the local three-particle correlation function $\g\equiv\braket{ \hat{\psi}^\dagger(x)^3 \hat{\psi}(x)^3}/n^3$, which compares the probabilities of having three particles at the same position for a correlated and an uncorrelated system. Here, $\hat{\psi}^\dagger$ and $\hat{\psi}$ are atomic field operators and $n$ is the density. The function $\g$ depends strongly on quantum statistics \cite{Burt1997,Hodgman2011} and temperature \cite{Kheruntsyan2003,Kormos2009}. For example, in 3D geometry, statistics change the value of $\g$ from zero for identical fermions to one for non-interacting classical particles and to six for thermal (non-condensed) bosons. For non-interacting bosons statistical bunching is suppressed in a Bose-Einstein condensate (BEC), for which $\g=1$. In addition, interactions also have a pronounced effect on $\g$: In a 3D BEC, quantum depletion due to quantum fluctuations reduces the condensate fraction by increasing the number of occupied single-particle modes. In this case, beyond-mean-field calculations \cite{Kagan1985} predict an increase of $\g$ proportional to the square root of the gas parameter $(n \scat^3)^{1/2}$, where $\scat$ is the 3D s-wave scattering length. This increase of $\g$ has never been seen experimentally and is in stark contrast to the behavior of 1D systems. In 1D geometry, bosons with repulsive interactions minimize their interaction energy by avoiding spatial overlap. For very strong repulsive interactions in the Tonks-Girardeau (TG) limit \cite{Girardeau1960,Kinoshita2004,Paredes2004,Haller2009,Cazalilla2011} a strong reduction of $\g$ with a $\gamma^{-6}$ scaling is predicted \cite{Gangardt2003}. Here, $\gamma$ is the dimensionless Lieb-Liniger parameter, which characterizes interactions in a homogeneous 1D system \cite{Cazalilla2011,Methods1}. Recently, $\g$ has been calculated all the way from the weakly to the strongly interacting 1D regime \cite{Cheianov2006}. Experimentally, Laburthe Tolra {\it et al.} \cite{Tolra2004} have observed a reduction of $\g$ by a factor of about $7(5)$ for a weakly interacting gas of Rb atoms with $\gamma=0.45$.

In this work we experimentally determine $\g$ in 3D and in 1D geometry using  a trapped ultracold gas of Cs atoms with tunable (repulsive) interactions. For a BEC in 3D geometry we find clear evidence for an increase of $\g$ with increasing interaction strength, in good agreement with the prediction of Ref.~\cite{Kagan1985}. In 1D, for which we can tune $\gamma$ from zero to above $100$ \cite{Haller2009}, we determine $\g$ in the crossover regime from weak (1D BEC regime) to strong interactions (TG regime). Here our data agrees well with the prediction of Ref.~\cite{Cheianov2006}. For strong interactions in the TG regime, our measurements show that $\g$ is suppressed by at least three orders of magnitude. For high densities and strong interactions, we observe a rather sudden increase of three-body losses after long hold times in the trap. Understanding the behavior of $\g$ at short and long times is an important step towards understanding integrability and thermalization in 1D systems~\cite{Kinoshita2006,Hofferberth2008}.

A three-body loss process \cite{Fedichev1996,Chin2010} consists of the collision of three  particles, the formation of a dimer, and the release of the dimer's binding energy typically sufficient to allow both, the dimer and the remaining particle, to escape from the trap. The loss, assuming negligible one- and two-body loss, is modeled by the rate equation $\dot{n} = -\alpha \K \g n^3$. Here, we have explicitly split the loss rate coefficient $\alpha \K \g$ into its three contributions. The parameter $\alpha=3$ describes a situation where exactly three particles are lost in each recombination event. In principle, secondary losses \cite{Zaccanti2009} could modify its value. However, in the following we will be interested in {\it relative} measurements of $\alpha\K\g$, which are only weakly dependent on the precise value of $\alpha$ \cite{Methods1}, allowing us to neglect a possible deviation of $\alpha$ from the value of $3$. The parameter $\K$ contains the effect of few-body physics on the loss process \cite{Chin2010}. It depends on the probability of dimer formation (a process that can be strongly enhanced near Efimov resonances \cite{Kraemer2006}) and generally varies strongly with $\scat$ \cite{Fedichev1996,Esry1999,Nielsen1999,Bedaque2000,Weber2003}. For $\scat$ much larger than the range of the scattering potential, $\K$ shows a generic $\scat^4$ scaling. Contributions of many-body physics are contained in the three-particle distribution function $\g n^3$. In what follows, we aim to measure $\g$ as a function of $\scat$ both in 3D and 1D geometry.

\begin{figure}[t]
\centering
\includegraphics[width=0.47\textwidth]{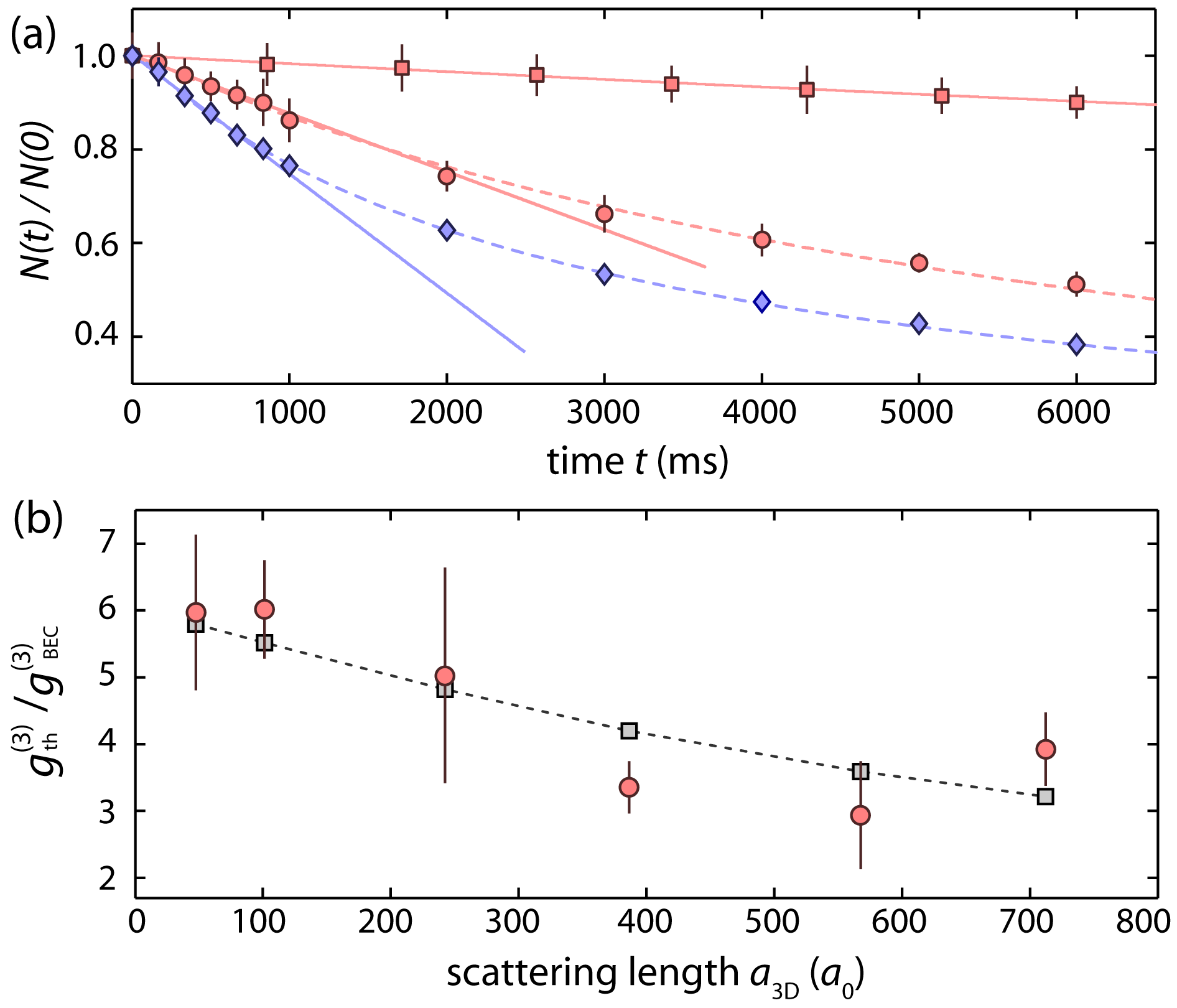}
\caption{(color online) (a) Relative atom number $N(t)/N(0)$ vs. hold time $t$ in 3D geometry: BEC (squares and circles) and thermal gas (diamonds) for $\scat=101(2)\ a_0$, $386(3)\ a_0$, and $386(3)\ a_0$, respectively. The dashed lines are fits to the data based on the loss equation (see text). The solid lines are linear fits that include the data from 100\% to 85\%. (b) The ratio of correlation functions $\g_\text{\tiny th}/\g_\text{\tiny BEC}$ as a function of $\scat$ (experimental data: circles; prediction \cite{Kagan1985}: squares). All error bars reflect the $1\sigma$ statistical uncertainty.
\label{fig:1}}
\end{figure}

\begin{figure}[t]
\centering
\includegraphics[width=0.47\textwidth]{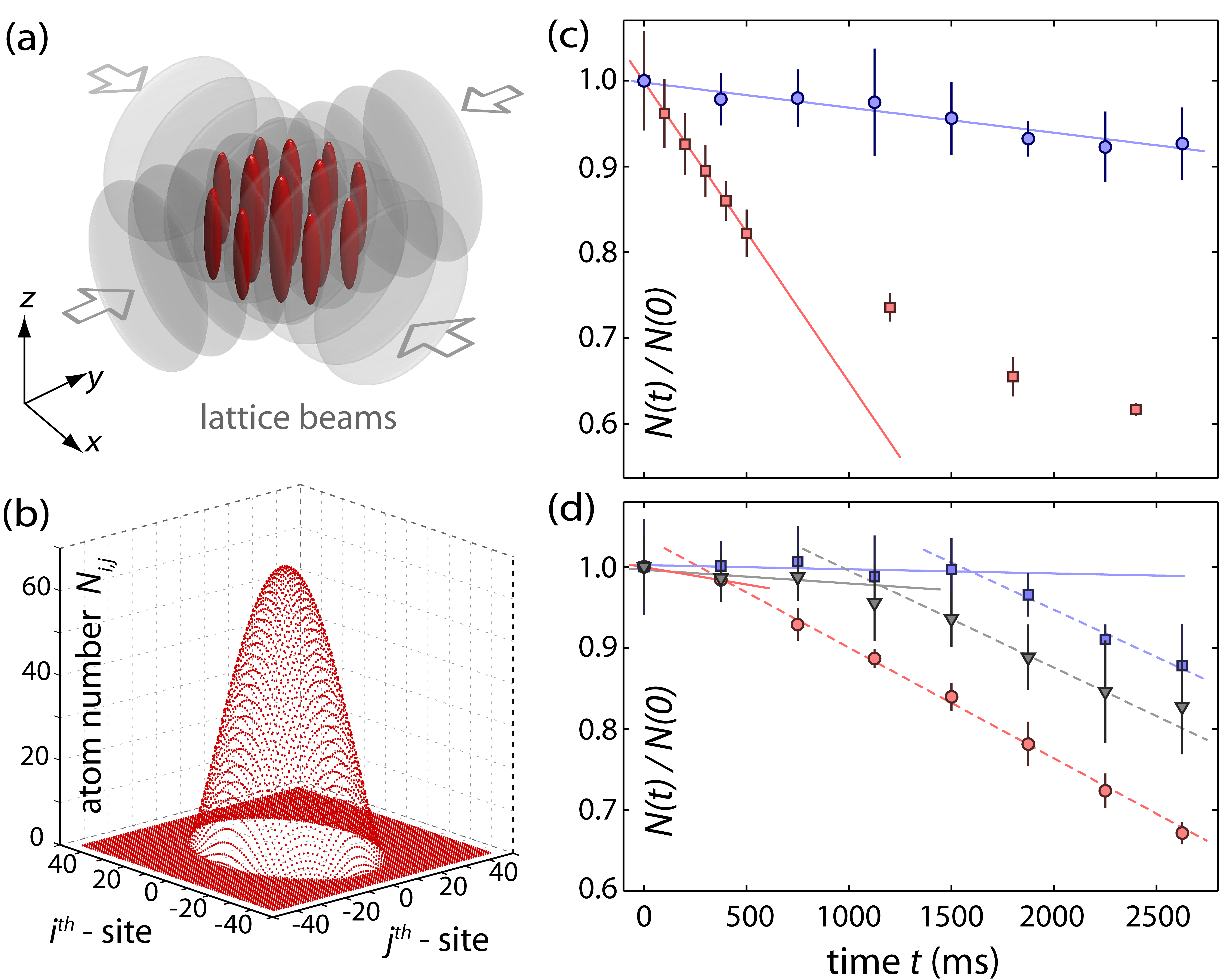}
\caption{(color online) (a) Sketch of the experimental setup: A 2D optical lattice
traps atoms in an array of 1D tubes. (b) Example of a computed atom number
distribution $N_{i,j}$ (see text). (c) The relative atom number $N(t)/N(0)$ as a
function of time $t$ in 1D geometry: squares and circles correspond to
$\scat = 23(1)\ a_0$ and $568(3)\ a_0$ with initial densities of
$4.5$ $\mu$m$^{-1}$ and $1.7$ $\mu$m$^{-1}$ at the center of the center tube, respectively.
The solid lines are linear fits to the initial slopes. (d) The relative atom number $N(t)/N(0)$ in
1D for fixed $\scat=568(3)\ a_0$ and for various values of $\gamma$ as the 1D density is
changed: $\gamma=12$ (circles), $\gamma=13$ (triangles), and $\gamma=14$ (squares). The
solid (dashed) lines are linear fits to the data points for short (large) times to guide
the eye. \label{fig:2}}

\end{figure}
We determine $\Kg$ from measurements of the decay of the total number of atoms $N(t)$ in our trap \cite{Burt1997,Weber2003}, which obeys the loss equation $\dot{N} = - 3\Kg \int n^3({\mathbf r}) d^3r $. Figures~\ref{fig:1}(a) and \ref{fig:2}(c) show typical atom number measurements for 3D and 1D geometry. The data in 3D geometry is well fit by solutions to the loss equation. The determination of $\Kg$ depends critically on an exact knowledge of the atomic density profile $n({\mathbf r})$. In particular, particle loss and loss-induced heating of the sample \cite{Weber2003} can modify the density profile in a non-trivial way. Also, on long time scales evaporative losses might start to play a role. To avoid these complications we restrict ourselves to short time intervals, during which not more than $15\%$ of the atoms are lost, and we determine the slope $\dot{N}(0)$ from a linear fit to the data. We determine $\int n^3({\mathbf r}) d^3r$ from a measurement of the total atom number $N$ and the trap frequencies $\omega_{x,y,z}$ using interaction dependent models for $n({\mathbf r})$ \cite{Methods1}. We find that the linear approximation underestimates $\Kg$ by approximately $12\%$, however, the data analysis is greatly simplified, especially in 1D. Finally, a comparative measurement of $\Kg$ allows us to eliminate $\K$, as explained below, and to determine $\g$ in 3D and 1D geometry.

{\it Correlation function in 3D:} We {measure $\K \g$ for both a non-condensed thermal sample  and a BEC as a function of $\scat$. For the thermal sample we start with typically $3.5 \times 10^5$ Cs atoms at a temperature of $T\!\approx\!200$ nK. The peak density is about $n_0 = 1\times10^{14}$ cm$^{-3}$. In the BEC \cite{Weber2003,Kraemer2004} we have about $9 \times 10^4$ Cs atoms without any detectable non-condensed fraction at about $n_0 = 5\times10^{13}$ cm$^{-3}$. We tune $\scat$ in the range from $50\ a_0$ to $800\ a_0$ by means of a broad magnetic Feshbach resonance \cite{Weber2003,Lange2009} ($a_0$ is Bohr's radius). The magnetic field gradient needed to levitate the atoms against gravity \cite{Weber2003} introduces a slight (less than $5\ a_0$) variation of $\scat$ across the samples. We determine $N$ by means of absorption imaging after a variable hold time $t$ and $50$ ms of expansion in the presence of the levitation field. We note that we do not observe the appearance of any non-condensed fraction in all measurements using the BEC. Figure \ref{fig:1}(b) displays the ratio $K^{(3)}_\stext{th} g^{(3)}_\stext{th} / (K^{(3)}_\stext{BEC} g^{(3)}_\stext{BEC}) = \g_\stext{th}/\g_\stext{BEC}$ determined from the thermal sample and the BEC as a function of $\scat$. Here we have made the reasonable assumption that $K^{(3)}$ is independent of the system's phase in 3D geometry, i.e. $K^{(3)}_\stext{th} = K^{(3)}_\stext{BEC}$. Our measurement shows that the ratio $\g_\stext{th}/\g_\stext{BEC}$ attains the expected value of $6$ for weak interactions \cite{Burt1997}, but then exhibits a pronounced decrease as $\scat$ is increased. For comparison, we plot the prediction of Ref.~\cite{Kagan1985}
\begin{eqnarray}
    \label{eq:corr} \g_\stext{th}/\g_\stext{BEC} = 6 / \left( 1+
    \frac{64}{\sqrt{\pi}}\sqrt{n_0 a^3_\stext{3D}} \right).
\end{eqnarray}
We note that the density $n_0$ enters into this equation as a measured quantity. In general,  we find good agreement between the experimental and the theoretical result, establishing our measurement as a clear demonstration of beyond mean-field effects on $\g$ in 3D bosonic quantum gases.

{\it Correlation function in 1D:} Figure \ref{fig:2} (a) illustrates our experimental setup to generate an array of 1D systems. We load a BEC of typically $8 \times 10^4$ atoms within $400$ ms into approximately 5000 vertically (z-direction) oriented tubes that are formed by two horizontally propagating, retro-reflected lattice laser beams. Each tube with index $(i,j)$ in the $x$-$y$-plane has a transversal trapping frequency of $\omega_\perp = 2\pi \times 12.2(5)$\ kHz and an aspect ratio $\omega_\perp/\omega_z$ of approximately $800$. The transversal motion of the atoms in the tubes is effectively frozen out as kinetic and interaction energy are much smaller than $\hbar \omega_\perp$. We adjust $\scat$ in $100$ ms to its final value. After time $t$ we turn off the lattice potential and determine the total atom number $N(t)$ by absorption imaging in a time-of-flight measurement. In order to determine $\g_\stext{1D}$ we calculate the ratio $\K_\stext{1D}\g_\stext{1D}/(\K_\stext{3D}\g_\stext{3D})=\g_\stext{1D}/\g_\stext{3D}$. Here, it is not obvious that few-body physics is not affected by the confinement and that hence $\K_\stext{1D}$ and $\K_\stext{3D}$ cancel each other. Nevertheless, it is reasonable to assume that $\K$ is not significantly changed by the confinement as long as the confinement length $a_\perp  = \sqrt{\hbar/(m \omega_\perp)}$ is larger than the extent of the dimer produced in the recombination event and the range of the scattering process, which are both of order of $\scat$. Here, $m$ is the atom mass. We choose a moderately deep lattice potential with $a_\perp\approx 1500\ a_0$ and restrict $\scat$ to $\scat \lesssim 800 \; a_0$. In particular, we avoid the confinement-induced resonance condition $\scat \approx a_\perp$
\cite{Haller2009,Haller2010}.

The main difficulty in the determination of $\K_\stext{1D}\g_\stext{1D}$ comes from the fact that the initial atom number of the tubes varies across the lattice as a result of the harmonic confinement. We choose to always load the lattice in a regime of weak repulsive interactions such that almost all 1D samples are initially in the 1D Thomas-Fermi (TF) regime \cite{Menotti2002}. The local chemical potentials $\mu_{i,j}$, the total atom number $N$, and the chemical potential $\mu$ are then unambiguously related, and we can directly calculate the initial occupation number $N_{i,j}$ for each tube $(i,j)$ (\cite{Methods1} and Fig.~\ref{fig:2}(b)). The variation in $N_{i,j}$ results in a considerable variation in the type of density profile for each of the 1D systems after the strength of interactions is increased to the desired value: Some tubes remain in the 1D TF regime, while others are now in the TG regime. For tubes that are in the weakly interacting regime we determine the 1D density $n_\stext{1D}$ numerically by solving the 1D Gross-Pitaevskii equation. For the TG regime the density profiles are determined following Ref.~\cite{Menotti2002}. In general, we find good agreement when we compare the numerical results to integrated density distributions from in-situ absorption images. For the interaction parameter $\gamma$ we take a mean value that is calculated as an average over all local $\gamma_{i,j}$ at the center of each tube $(i,j)$ weighted by $N_{i,j}$ \cite{Methods1}.

As before we determine $\K_\stext{1D}\g_\stext{1D}$ from the initial slope of the loss curve as shown in Fig.~\ref{fig:2}(c). In Fig.~\ref{fig:3}(a) we compare the data that we obtain in 1D geometry to our data for $\K_\stext{3D}\g_\stext{3D}$ for a 3D-BEC as we vary $\scat$. We note that the BEC data is in good agreement with previous three-body loss data on thermal samples when one takes into account the combinatorial factor $3!=6$ \cite{Weber2003,Kraemer2006}. In particular, the 3D data follows the universal scaling law $\K \sim \scat^4$ for sufficiently large $\scat$ \cite{Fedichev1996,Esry1999,Nielsen1999,Bedaque2000,Weber2003}. We exclude data points affected by the presence of a narrow Feshbach resonance in the vicinity of $\scat = 150 a_0$ \cite{Mark2007}. Note that in the range from $\scat \approx 10 a_0$ to $\scat \approx 850 a_0$ three-body losses in 3D increase by nearly 3 orders of magnitude. This behavior is in stark contrast to the measurements in 1D. In 1D, we observe a {\it reduction} of $\Kg$ by approximately a factor of $2$ upon increasing $\scat$ over the same range of values. In fact, for $\scat \ge 200 a_0$ our measurement only gives an upper bound on $\K_\stext{1D}\g_\stext{1D}$ as losses become so small that we have difficulty in determining $\dot{N}(0)$. Note that tunneling between tubes (on a timescale of 1 s for the parameters of our lattice) sets an upper bound for the timescale for which the tubes can be considered to be independent and hence fully in the 1D regime.

\begin{figure}[t]
\centering
\includegraphics[width=0.47\textwidth]{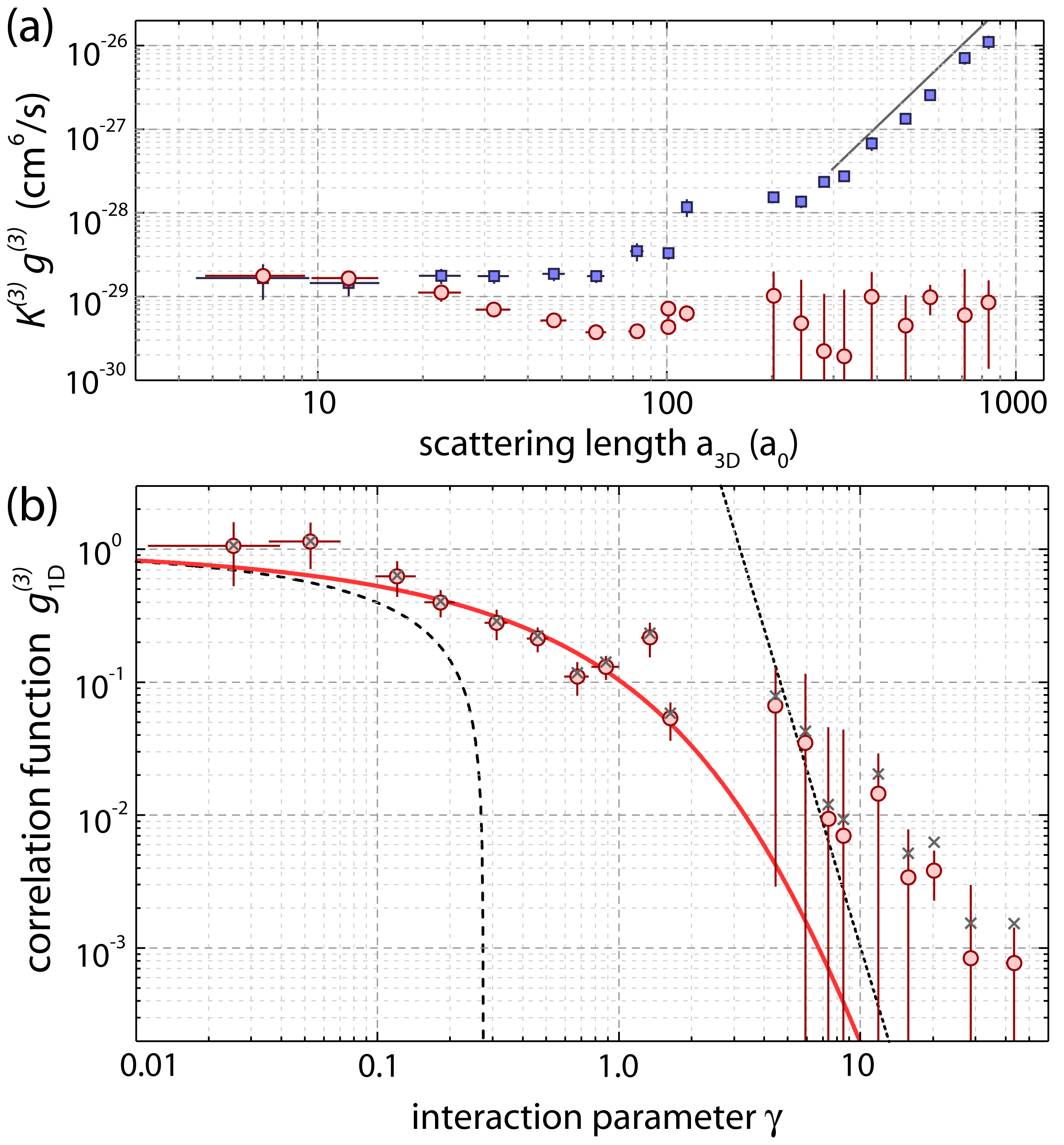}
\caption{\label{fig:3} (color online) (a) Three-body loss coefficient $K^{(3)}g^{(3)}$ vs. $\scat$
for a quantum degenerate gas in 3D (squares) and in 1D (circles). The line gives
the $K^{(3)} = C \hbar \scat^4/m$ scaling in the universal regime in 3D with
$C=67.9$ \cite{Fedichev1996,Esry1999,Nielsen1999,Bedaque2000,Weber2003}. The error bars of $K^{(3)}g^{(3)}$ reflect the $1\sigma$ statistical uncertainty of the linear fit.
(b) The measured correlation function $g^{(3)}_\stext{1D}$ vs. $\gamma$ in 1D geometry (circles).
The crosses indicate the values for $g^{(3)}_\stext{1D}$ corrected for the
variations of $g^{(3)}_\stext{3D}$ with $\scat$ as in Eq.~\eqref{eq:corr}
(see text). The dashed and dotted lines are approximate analytical solutions
for $\gamma\ll1$ and $\gamma\gg1$ from Ref.~\cite{Gangardt2003}. The solid
line is the prediction from Ref.~\cite{Cheianov2006}.}
\end{figure}

In Fig.~\ref{fig:3}(b) we plot $\K_\stext{1D}\g_\stext{1D} /(\K_\stext{3D}\g_\stext{3D}) \approx \g_\stext{1D} $ as a function of $\gamma$. A striking decrease by 3 orders of magnitude from the value $1$ at $\gamma\approx 0.03$ to $10^{-3}$ at $\gamma \approx 50$ can be seen. We compare this result to the predictions based on the Lieb-Liniger model of interacting bosons in 1D: In the weakly interacting Gross-Pitaevskii regime ($\gamma \ll 1$) the Bogoliubov approach yields $\g(\gamma)\simeq 1 - 6 \sqrt{\gamma}/\pi$, while in the TG regime, $\gamma \gg 1$, $\g$ can be expressed through derivatives of the three-body correlation function of free fermions, giving $\g  = 16 \pi^6 / (15 \gamma^6)$ \cite{Gangardt2003}. Cheianov {\it et al.}~\cite{Cheianov2006} have recently calculated numerically $\g$ for all strengths of interactions within the Lieb-Liniger model, providing an interpolation between the weakly and strongly interacting limits (red continuous line in Fig.~\ref{fig:3}(b)). We find very good agreement between the result of our experiment and the theory that is valid for all strengths of interactions. This is the central result of this work.

Finally, for large values of $\scat$ and $n_0$, and for long hold times in 1D geometry, we find a surprisingly sudden increase of losses as shown in Fig.\ref{fig:2}(d), accompanied by a rapid increase for the expansion energy in the longitudinal direction (data not shown). The onset of increased losses shifts to later times with decreased density in the tubes, i.e. increased $\gamma$, and it is rather sensitive to the precise value of $\gamma$. We believe that the 1D tubes suffer from a recombination-heating induced breakdown of correlations: For sufficiently large values of $\scat$ the binding energy of the weakly bound dimer produced in the recombination process becomes comparable to the trap depth (here $h \times 45$ kHz). This leads to a positive feedback cycle in the many-body system in which three-body losses lead to an increase of temperature \cite{Weber2003} and thus of $\g$ \cite{Kheruntsyan2003}, which in turn increases three-body losses.

In summary, we have measured the local value $\g$ for the three-particle correlation function for quantum degenerate gases in 3D and 1D. In 3D, increasing interactions deplete the condensate and increase the value of $\g$ in accordance with beyond mean-field calculations. In 1D, we observe a strong suppression for $\g$ by 3 orders of magnitude as the TG regime is entered. The accompanying suppression of three-body losses is crucial to the study of strongly interacting matter in and out of equilibrium in 1D \cite{Kinoshita2006,Hofferberth2008,Haller2009,Haller2010a}.

We thank R. Grimm for generous support. We gratefully acknowledge funding by the Austrian Science Fund (FWF) within project I153-N16 and within the framework of the European Science Foundation (ESF) EuroQUASAR collective research project QuDeGPM. GP acknowledges funding from the EU through NAME-QUAM and AQUTE.

\bibliographystyle{apsrev}

\begin{references}

\bibitem{Cazalilla2011} M.A. Cazalilla {\it et al.}, arXiv:1101.5337.

\bibitem{Bloch2008} I. Bloch, J. Dalibard, and W. Zwerger, Rev. Mod. Phys. \textbf{80}, 885 (2008).

\bibitem{Chin2010} C. Chin, R. Grimm, P. Julienne, and E. Tiesinga, Rev. Mod. Phys. {\bf 82}, 1225 (2010).

\bibitem{Kinoshita2005} T. Kinoshita, T. Wenger, and D. Weiss, Phys. Rev. Lett. {\bf 95}, 190406 (2005).

\bibitem{Burt1997} E. Burt {\it et al.}, Phys. Rev. Lett. {\bf 79}, 337 (1997).

\bibitem{Tolra2004} B. Laburthe Tolra, {\it et al.}, Phys. Rev. Lett. {\bf 92}, 190401 (2004).

\bibitem{Yasuda1996}M. Yasuda and F. Shimizu, Phys. Rev. Lett. {\bf 77}, 3090 (1996).

\bibitem{Oettl2005} A. \"Ottl, S. Ritter, M. K\"ohl, and T. Esslinger, Phys. Rev. Lett. {\bf 95}, 090404 (2005).

\bibitem{Schellekens2005} M. Schellekens {\it et al.}, Science {\bf 310}, 648 (2005).

\bibitem{Jeltes2007} T. Jeltes {\it et al.}, Nature {\bf 445}, 402 (2007).

\bibitem{Greiner2005} M. Greiner, C. A. Regal, J. T. Stewart, and D. S. Jin, Phys. Rev. Lett. {\bf 94}, 110401 (2005).

\bibitem{Foelling2005} S. F\"olling {\it et al.}, Nature {\bf 434}, 481 (2005).

\bibitem{Rom2006} T. Rom {\it et al.}, Nature {\bf 444}, 733(2006).

\bibitem{Jacqmin2011} T. Jacqmin {\it et al.}, Phys. Rev. Lett. {\bf 106}, 230405 (2011).

\bibitem{Armijo2010} J. Armijo, T. Jacqmin, K. Kheruntsyan, and I. Bouchoule, Phys. Rev. Lett. \textbf{105}, 230402 (2010).

\bibitem{Hodgman2011} S.S. Hodgman {\it et al.}, Science \textbf{331}, 1046 (2011).

\bibitem{Kheruntsyan2003} K. Kheruntsyan,  D. Gangardt,  P. Drummond, and G. Shlyapnikov, Phys. Rev. Lett {\bf 91}, 040403 (2003).

\bibitem{Kormos2009} M. Kormos, G. Mussardo, and A. Trombettoni, Phys. Rev. Lett. {\bf 103}, 210404 (2009).

\bibitem{Kagan1985} Yu. Kagan, B.V. Svistunov, and G.V. Shlyapnikov, Pis'ma Zh. Eksp. Teor. Fiz. {\bf 42}, 169-172 (1985).

\bibitem{Girardeau1960} M. Girardeau, J. Math. Phys. {\bf{1}}, 516 (1960).

\bibitem{Kinoshita2004} T. Kinoshita, T. Wenger, and D. S. Weiss, Science {\bf 305}, 1125 (2004).

\bibitem{Paredes2004} B. Paredes {\it et al.}, Nature \textbf{429}, 277 (2004).

\bibitem{Haller2009} E. Haller {\it et al.}, Science \textbf{325}, 1224 (2009).

\bibitem{Gangardt2003} D. Gangardt and G. Shlyapnikov, Phys. Rev. Lett. {\bf 90}, 010401 (2003).

\bibitem{Methods1} See Supplemental Material at (URL will be inserted by publisher).

\bibitem{Cheianov2006} V.V. Cheianov, H. Smith, and M.B. Zvonarev, JSTAT {\bf 8}, P08015 (2006).

\bibitem{Kinoshita2006} T. Kinoshita, T. Wenger, and D.S. Weiss, Nature {\bf 440}, 900 (2006).

\bibitem{Hofferberth2008} S. Hofferberth {\it et al.}, Nature Phys. {\bf 4}, 489 (2008).

\bibitem{Fedichev1996} P.O. Fedichev, M.W. Reynolds, and G.V Shlyapnikov, Phys. Rev. Lett. {\bf 77}, 2921 (1996).

\bibitem{Zaccanti2009} M. Zaccanti {\it et al.}, Nature. Phys. {\bf 5}, 586 (2009).

\bibitem{Kraemer2006} T. Kraemer {\it et al.}, Nature {\bf 440}, 315 (2006).

\bibitem{Esry1999} B. D. Esry, C. H. Greene, and J. P. Burke, Phys. Rev. Lett. {\bf 83}, 1751 (1999).

\bibitem{Nielsen1999} E. Nielsen and J.H. Macek, Phys. Rev. Lett. {\bf 83}, 1566 (1999).

\bibitem{Bedaque2000} P.F. Bedaque, E. Braaten, and H.-W. Hammer, Phys. Rev. Lett. {\bf 85}, 908 (2000).

\bibitem{Weber2003} T. Weber {\it et al.}, Phys. Rev. Lett. {\bf 91}, 123201 (2003).

\bibitem{Kraemer2004} T. Kraemer {\it et al.}, Appl. Phys. B {\bf 79}, 1013 (2004).

\bibitem{Lange2009} A.D. Lange {\it et al.}, {\it Phys. Rev. A} {\bf 79}, 013622 (2009).

\bibitem{Haller2010} E. Haller {\it et al.}, Phys. Rev. Lett. \textbf{104}, 153203 (2010).

\bibitem{Menotti2002} C. Menotti and S. Stringari, Phys. Rev. A {\bf 66}, 043610 (2002).

\bibitem{Mark2007} M. Mark {\it et al.}, Phys. Rev. A {\bf 76}, 042514 (2007).

\bibitem{Haller2010a} E. Haller {\it et al.}, Nature {\bf 466}, 597 (2010).

\end{references}

\begin{references}
    \bibitem{Petrov2000M}  D.S. Petrov, M. Holzmann, and G.V. Shlyapnikov, Phys. Rev. Lett. {\bf 84}, 2551 (2000).
    \bibitem{Olshanii1998M} M. Olshanii, Phys. Rev. Lett. {\bf 81}, 938 (1998).
    \bibitem{Dunjko2001M} V. Dunjko, V. Lorent, and M. Olshanii, Phys. Rev. Lett. {\bf 86}, 5413 (2001).
    \bibitem{Menotti2002M} C. Menotti and S. Stringari, Phys. Rev. A {\bf 66}, 043610 (2002).
    \bibitem{Schuster2001M} J. Schuster {\it et al.}, Phys. Rev. Lett. {\bf 87}, 170404 (2001).
    \bibitem{Zaccanti2009M} M. Zaccanti {\it et al.}, Nature. Phys. {\bf 5}, 586 (2009).
\end{references}


\clearpage
\newpage

\section{Supplementary material}

\subsection{Trap parameters}
In 3D geometry, we measure the atom loss in a crossed beam dipole trap with one horizontal and one vertical laser beam. The horizontal trap-frequencies $\omega_{x,y}$ and the vertical trap-frequency $\omega_z$ vary for the different measurements. The data sets in Fig.\ref{fig:1}(a) are taken with trap frequencies $\omega_{x,y,z} = 2\pi \times (29(1), 80(2), 74(1))$ Hz for thermal atoms and with $\omega_{x,y,z} = 2\pi \times (11.8(1), 17.9(3), 13.5(1))$ Hz for a BEC. The data sets in Fig.\ref{fig:3}(a) are taken at trap frequencies of $\omega_{x,y,z} = 2\pi \times (10.5(8),17.4(1),13.9(1))$ Hz for a BEC.

In 1D geometry, we use a crossed dipole trap in addition to the 2D optical lattice potential to adjust the atom number distribution over the tubes. We choose two settings with global trap frequencies $\omega_{x,y,z}=2\pi\times(9.7(2),11.4(2),14.5(1))$ Hz and $2\pi\times(13.1(2),17.7(2),17.5(2))$ Hz.

\subsection{Atom number distribution over the tubes}
We calculate the initial occupation number $N_{i,j}$ for tube $(i,j)$ from the global chemical potential $\mu$. For weak repulsive interactions during the loading process almost all tubes are in the 1D Thomas-Fermi (TF) regime with a local chemical potential $\mu_{i,j}$ at the center of each tube
\[ \mu_{i,j} = \mu - \frac{1}{2}m (\lambda/2)^2 (\omega_x^2 i^2+\omega_y^2 j^2),\]
where m is the atomic mass and $\lambda=1064.5\ $nm is the wavelength of lattice light. We calculate $\mu$ from the condition $N = \sum_{i,j} N_{i,j}(\mu)$ with the $N_{i,j}$ given by \cite{Petrov2000M}
\[ \mu_{i,j} = \left( \frac{3 N_{i,j}}{4\sqrt{2}} g_\stext{1D}\omega_z \sqrt{m} \right)^{2/3},\]
where $g_\stext{1D}$ is the 1D coupling parameter \cite{Olshanii1998M}
\begin{eqnarray*}
    g_\stext{1D} = 2\hbar \omega_\perp \scat \left(1-
    1.0326\frac{\scat}{a_\perp}\right)^{-1}.
\end{eqnarray*}

\subsection{Determination of $\gamma$}
We determine the mean interaction parameter $\gamma$ from the local parameters $\gamma_{i,j}$ at the center of each tube $(i,j)$
\begin{eqnarray*}
 \gamma_{i,j} = \frac{ m g_\stext{1D}}{\hbar^2 n^\text{1D}_{i,j}}, \qquad
 \gamma = \frac{1}{N} \sum_{i,j} N_{i,j}\gamma_{i,j}.
\end{eqnarray*}
Here, $n^\text{1D}_{i,j}$ is the 1D density at the center of the tube $(i,j)$. Note that this gives a lower estimate for $\gamma$. Averaging $\gamma$ over the density profile along each tube gives a slightly larger $\gamma$ by a factor 1.5 for a 1D TF density profile and a factor 1.27 for a TG density profile.

\begin{figure}
\centering
\includegraphics[width=0.45\textwidth]{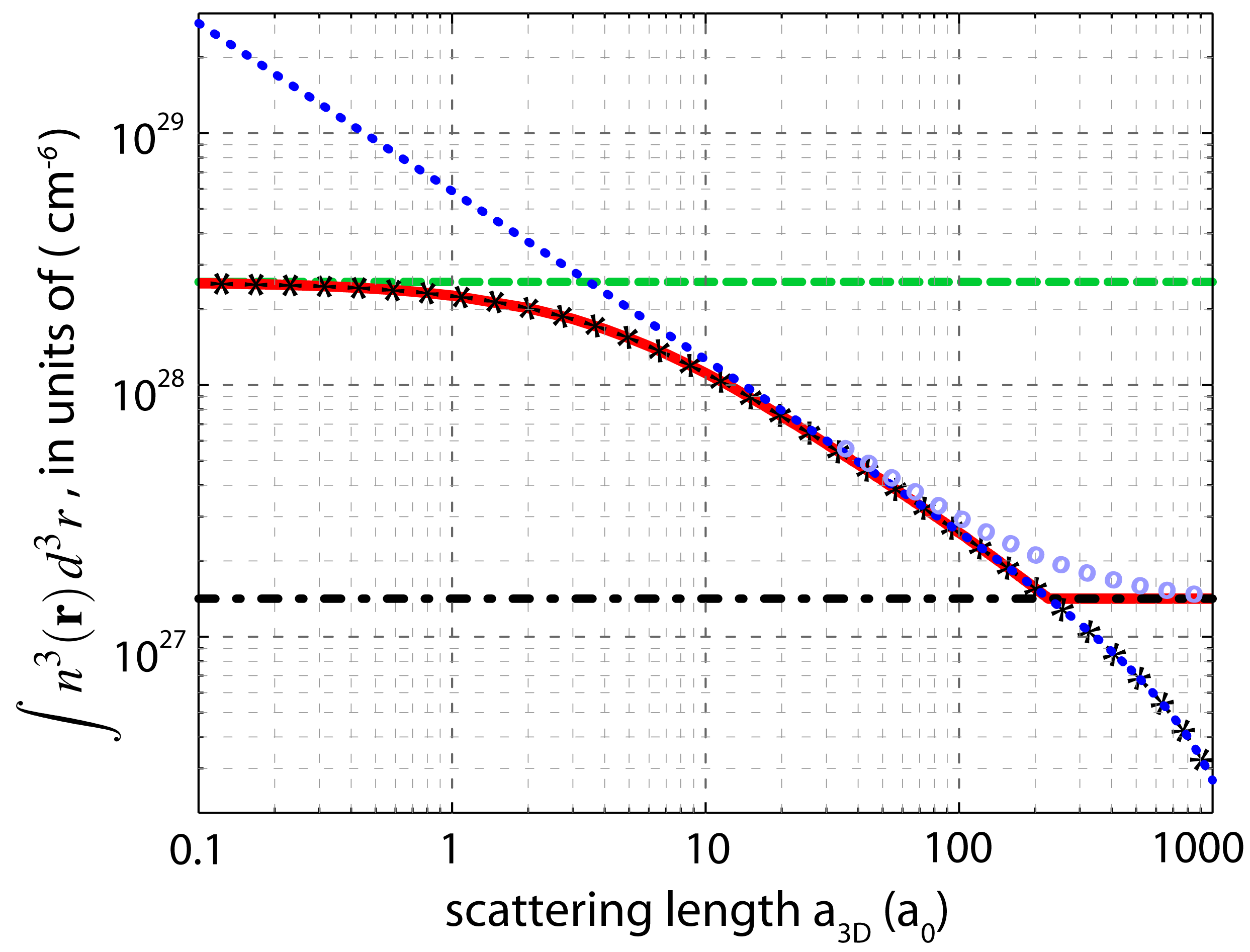}
\caption{(color online) The quantity $\int n_{i,j}^3({\mathbf r}) d^3r$ for an occupation
number $N_{i,j}=15$ as a function of the scattering length $\scat$ for the
given trap parameters of our experiment. The various
curves correspond to the different approximations: gaussian solution (green dashed line),
TF solution (blue dotted line), TG solution (black dash-dotted line), numerically solved GP-equation result
(black stars), and Lieb-Liniger solution with local density
approximation (blue circles). For the data analysis the continuous red line is used. \label{sfig:2}}
\end{figure}

\subsection{Density profiles}

The density profiles for the individual tubes with index $(i,j)$ depend strongly on the strength of interactions and the occupation number $N_{i,j}$.
Fig.~\ref{sfig:2}} compares the results for the integrated density profiles $\int n_{i,j}^3({\mathbf r}) d^3r$ using the different approximations to calculate the profile (gaussian, TF, TG, numerically solved GP-equation, and Lieb-Liniger solution within the local density approximation \cite{Dunjko2001M}) for the specific case of $N_{i,j}=15$. For our analysis of the experimental data we use the GP result for weak interactions and the TG result \cite{Menotti2002M} for strong interactions (continuous red line).

\subsection{Secondary loss processes}

Here we estimate the deviation $\Delta \alpha$ from $\alpha=3$ in the rate equation $\dot{n} = -\alpha \K \g n^3$ due to secondary loss processes \cite{Schuster2001M,Zaccanti2009M}. Within a simple simulation, we determine an upper bound for the correction to the data of the 3D loss experiment of Fig.~\ref{fig:1}(b) and show that secondary loss processes cannot explain our results for $\g_\stext{th}/\g_\stext{BEC}$. In fact for our experimental trap parameters and atom numbers secondary processes would result in an {\it increase} of $\g_\stext{th}/\g_\stext{BEC}$ with increasing interaction strength, in contradiction with the observed behavior.

\begin{figure}
\centering
\includegraphics[width=0.45\textwidth]{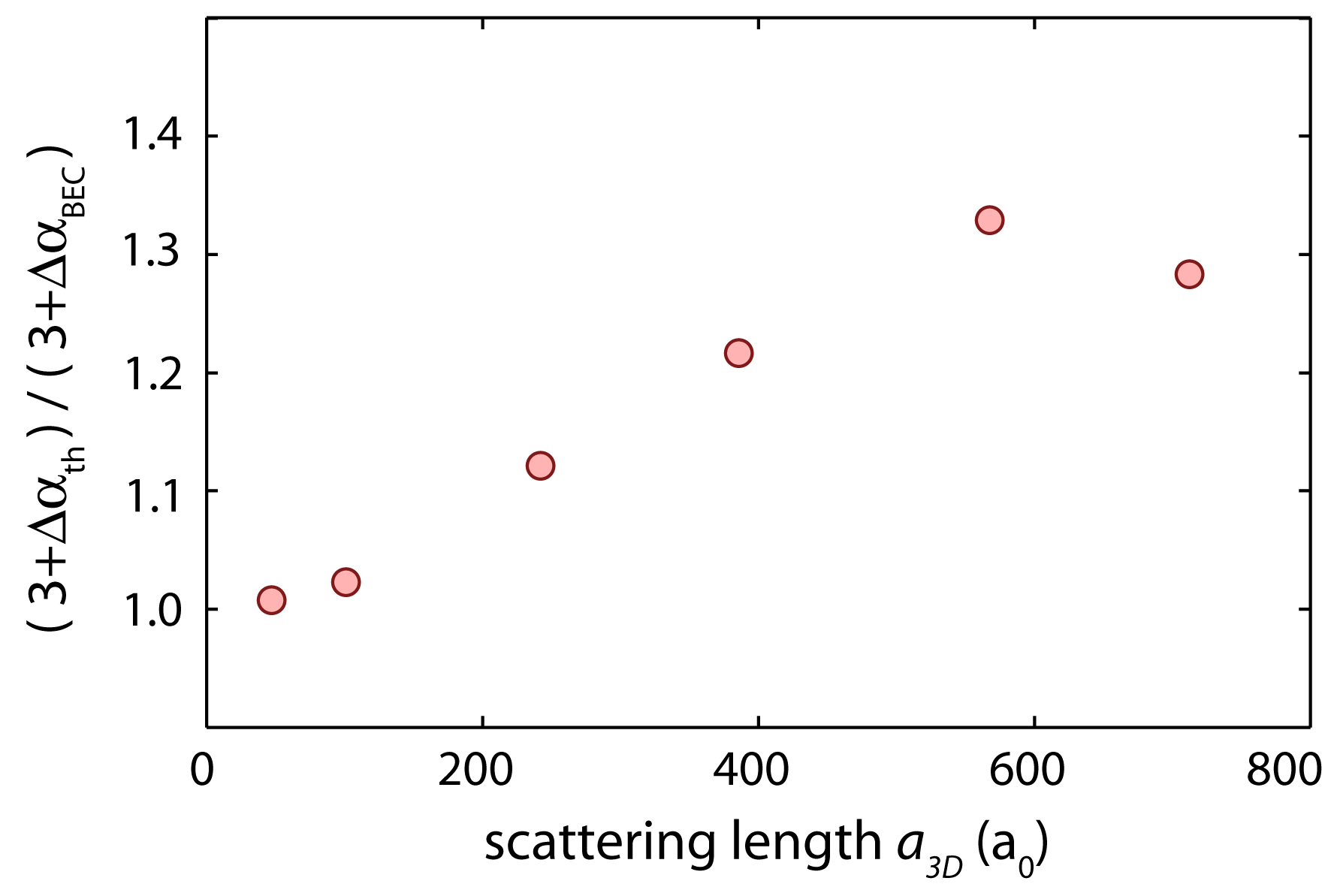}
\caption{(color online) Upper estimate for the corrections due to secondary loss. For details see text.\label{sfig:1}}
\end{figure}

A secondary collision is caused by the collision of the dimer and/or free atom from a three-body recombination process with other atoms while leaving the trap, triggering additional losses. We estimate the average number of secondary collisions in our experiment by determining numerically the collisional opacity $\braket{nl}\sigma$ for the products of a three-body recombination event. Here $\braket{nl}$ is the average column density, with $l$ the distance covered by a (randomly chosen) atom leaving the trap, and $\sigma$ is the scattering cross section. For the atom-atom cross section we use the formula $\sigma = 8\pi \scat^2/(1+k^2\scat^2)$, where $\hbar k$ is the momentum of the free atom gained in the recombination event. For the atom-dimer collision, we use a similar expression for the cross section, with the momentum of the dimer and a scattering length $2\scat$. For our experimental parameters, we then determine the total number of atoms lost due to secondary processes for a thermal sample, $\Delta \alpha_\stext{th}$, and a BEC, $\Delta \alpha_\stext{BEC}$. Figure~\ref{sfig:1} shows that the ratio $(3+\Delta \alpha_\stext{th})/(3+\Delta \alpha_\stext{BEC})$ {\it increases} with increasing $\scat$ by about 30 percent over the experimentally accessible range of $\scat$. These results imply that for our experimental parameters secondary loss processes would result in an increase of the ratio $\g_\stext{th}/\g_\stext{BEC}$ in Fig.~\ref{fig:1}(b), in contrast to the measured data. Thus, within the present model, this rules out secondary loss processes as the cause of the effects shown in the present work.

\end{document}